\documentclass{aa}  

\usepackage{graphicx}
\usepackage{txfonts}
\usepackage{lipsum}
\usepackage{subcaption}         
\newcommand{\atanTwo}{\operatorname{atan2}}
                                
\usepackage{lscape}             
\usepackage{placeins}           
\usepackage{xcolor}
\usepackage{float}
\usepackage{multirow}
\usepackage{hyperref}
\usepackage{comment}
\usepackage{ulem}

\begin{document}

   \title{StreakMind: AI detection and analysis of satellite streaks in astronomical images with automated database integration}

   \author{Rafael Carrillo\inst{1,2,3}, René Duffard\inst{3},  Pablo García-Martín\inst{4}
        , Javier Romero\inst{5}, Nicolás Morales\inst{3} \and Luis Gonçalves\inst{6}
        }

   \institute{
    Real Instituto y Observatorio de la Armada (ROA), Plaza de las Marinas s/n, 11100 San Fernando (Cádiz), Spain
    \and
    Universidad de Granada, Departamento de Física Teórica y del Cosmos, 18071 Granada, Spain
    \and
    Instituto de Astrofísica de Andalucía -- CSIC, Apdo. 3004, 18080 Granada, Spain
    \and
     Safran, 171 Bd de Valmy, 92700 Colombes, France
    \and
    Universidad de Cádiz, Departamento de Matemáticas, 11510 Puerto Real (Cádiz), Spain
    \and
    CFisUC, Departamento de F\'{\i}sica, Universidade de Coimbra, 3004-516 Coimbra, Portugal
    }

   \date{Received 22 December 2025 / Accepted 4 March 2026}

  \abstract
    {Artificial satellites and space debris are increasingly contaminating astronomical images, affecting scientific surveys and producing large volumes of streaked exposures. Manual inspection is no longer feasible at scale, and reliable identification and characterisation of streaks has become essential for both the quality control of data and the monitoring of objects in Earth orbit.}
    {We present StreakMind, an automated pipeline designed to detect near-Earth objects (NEOs) and satellite streaks in astronomical images, characterise their geometry, and cross-identify them with known orbital objects. The system integrates all inference results into a structured database suitable for large surveys.}
    {A YOLO-OBB model was trained on a hybrid manual–synthetic dataset of 2335 images and used to detect streaks in processed FITS frames. Geometric refinement, inter-frame association, satellite cross-identification, and Gaussian-based confidence scoring were then applied to produce final identifications, which were stored in a normalised relational database. In this work, images acquired at La Sagra Observatory (L98) with a Celestron C14+Fastar telescope were used to develop and test automated streak detection and characterisation methods.}
    {On the test set, the model achieved a precision of 94\% and a recall of 97\%. It reliably detected faint streaks, delivered consistent geometric reconstructions across the dataset, and performed robust satellite cross-identification. The Gaussian-based confidence scoring provided stable identification probabilities across consecutive frames.}
    {StreakMind demonstrates strong potential for large-scale automated analyses of linear streaks produced by both NEOs and artificial satellites in ground-based astronomical images. The pipeline offers high detection reliability, robust geometric reconstruction, and reproducible satellite cross-identification within a fully integrated end-to-end framework.}


   \maketitle
   
\section{Introduction}
The advent of wide-field high-cadence surveys has revolutionised the detection of near-Earth objects (NEOs), defined as asteroids with an orbit within 1.3 AU from the Sun \citep{Bottke_02}, shifting NEO detection and Solar System survey science into an era of big data. Modern facilities now monitor the sky with unprecedented speed, generating a volume of imagery that renders manual inspection unfeasible. While the discovery of these Solar System objects remains a primary scientific goal and the cornerstone for planetary defense, this continuous surveillance is equally critical for space situational awareness, especially as it relates to the monitoring of the growing population of artificial satellites and debris essential for maintaining orbital operations. Consequently, detection efforts must now navigate a crowded dynamic environment and distinguish faint NEOs from the dense traffic of Earth's growing in-orbit technological ecosystem.

Several methods can be used to find fast moving objects close to Earth in astronomical images. One approach uses a type of algorithm that relies on the identification of linear features and is based on methods such as the Hough transform \citep{Billot_24, Goncalves_26} and the Radon transform \citep{Stark_2022}. Alternatively, synthetic tracking has emerged as a powerful technique for discovering faint NEOs. By digitally shifting and stacking multiple short exposures along hypothetical velocity vectors, this method dramatically increases the signal-to-noise ratio (S/N) of moving targets that are otherwise undetectable in individual frames \citep{Stanescu_25, Vaduvescu_25, Zhai_24}. In this study, we describe a machine learning tool we designed to detect moving objects in astronomical images. 

Some previous works have highlighted the advantages of applying machine learning techniques to transient detection in astronomical images. From space, these methods have been employed to detect streaks in Hubble Space Telescope observations \citep{Kruk_22}, in the CHEOPS mission archive (García-Martín et al. in prep.), and in simulated imagery for the Euclid mission \citep{Pontinen_23,Lieu_19}. From the ground, the Zwicky Transient Facility has a long history of integrating deep learning to automate the identification of fast-moving objects. \citet{Duev_19} introduced a classifier operating on difference images, reducing human scanning time. Most recently, \citet{Irureta_25} advanced this effort by implementing an object segmentation approach trained on a mix of real and synthetic data, demonstrating the ability to recover hundreds of valid streaks missed by human scanners. 

With this paper we present StreakMind, a comprehensive AI-based data pipeline designed for the automated detection and characterisation of streaks in ground-based astronomical images. The detection stage is generic and applicable to any linear moving feature; however, the identification component developed in this work focuses on artificial satellites and space debris through cross-matching against external ephemerides. All detections are integrated into a centralised database that supports cross-referencing with catalogues of artificial objects. The extension of the identification layer to NEOs is planned as future work. 

Automatically detecting and characterising faint linear streaks produced by artificial satellites and debris in large volumes of astronomical imaging data is a significant challenge, particularly when these artefacts must be distinguished from genuine Solar System objects and instrumental effects. We developed an end-to-end pipeline to (i) detect linear streaks in ground-based astronomical images, (ii) refine their geometry and associate detections across consecutive frames, (iii) standardise the resulting measurements into Minor Planet Center (MPC)-style records, and (iv) cross-identify candidate artificial objects using external ephemerides, integrating all outputs into a relational database suitable for large-scale analyses.

\section{Methods and data}

This section describes the data and methodological framework used to develop and evaluate the StreakMind pipeline. We first describe the detection pipeline and model architecture (Section~\ref{subsec:pipeline}). We then summarise the telescope instrumentation and the FITS dataset (Sections~\ref{subsec:telescope}--\ref{subsec:fits}), followed by the generation of synthetic streaks (Section~\ref{subsec:synthetic_data}). The subsequent subsection details the labelling procedure and the FITS-to-PNG conversion including oriented bounding boxes (OBBs; Section~\ref{subsec:labeling}). Finally, Section~\ref{subsec:stratification} outlines the stratified sampling strategy adopted for training, validation, and testing.

At the core of this workflow lies the automated detection of linear streaks in individual images. This task is addressed using a deep-learning–based object detector, whose architecture and configuration are described in the following subsection.

\subsection{Pipeline architecture}\label{subsec:pipeline}

You Only Look Once (YOLO) is a family of real-time object detection models that predict both the location and the category of objects in a single pass through a neural network. By combining object localisation and classification into a single-stage process, YOLO enables fast and efficient inference, making it suitable for time-critical applications. The YOLO11 version, introduced in 2024 \citep{yolo11_ultralytics}, incorporates several architectural improvements that enhance both detection accuracy and computational performance. The model maintains the traditional three-part structure consisting of a backbone, neck, and head. The backbone extracts visual features from the input image at different levels of detail. The neck then combines these features across multiple scales, allowing the model to detect both small and large objects more effectively. Finally, the head processes this information to produce the final predictions, including object positions and class labels. 
Further details on the architectural components and multitask capabilities of YOLO11 can be found in \citet{Khanam_24}.

For this work, we adopted a pretrained YOLO11 model configured for OBBs, pretrained on the DOTAv1.0 dataset \citep{Xia_2018}. For the specific purpose of streak detection, the network was subsequently trained on an augmented dataset comprising both real and synthetically generated astronomical images, as described in detail in Section~\ref{subsec:synthetic_data}. Training was carried out on a cloud-based NVIDIA A100 GPU, using eight worker threads to accelerate data loading.

\subsection{Telescope and instrumentation}\label{subsec:telescope}

Observations were conducted at the La Sagra Observatory, located at latitude 37.982402°N, longitude 2.565290°W, at an altitude of 1513.654841 metres (MPC code L98). The telescope used is a Celestron C14+Fastar f/2.1 reflector, designated internally as Tetra1, mounted on a German Paramount mount. The instrument has an aperture of 356 mm and a focal length of 712 mm. It is equipped with an SBIG ST-10 3 CCD camera, providing a pixel scale of 4.12 arcsec/pixel and a field of view of 74.9 × 50.5 arcminutes. The native pixel size of the detector is 6.8~$\mu\text{m}$, and images were acquired with 2×2 binning. This binning mode was adopted to reduce data volume and facilitate nightly data transfer, as the observations were conducted within the framework of an external collaboration requiring the download of the full datasets after each observing session. The substantial pixel scale results from the integration of Fastar optics at the primary focus, reducing the focal ratio from f/10 to f/2. While this configuration increases the image scale, it successfully provides the wider field of view required for the instrument's primary scientific applications. No physical photometric filter was employed during acquisition.

\subsection{FITS dataset description}\label{subsec:fits}

To train and evaluate the StreakMind pipeline, we relied on a combination of real astronomical images and synthetically generated data. This subsection describes the real FITS image dataset and motivates the need for the inclusion of synthetic streaks, whose generation and properties are detailed separately in Section~\ref{subsec:synthetic_data}.

The real dataset consists of a total of 2055 FITS images, originally acquired for astrometric and photometric studies of asteroids, with the target objects generally located near the centre of the field of view. Table~\ref{table:1} summarises the observing dates between April and June 2019, indicating the specific asteroid targeted each night and the number of images obtained.

\begin{table}[H]
\caption{Distribution of FITS images included in the dataset.}
\label{table:1}
\centering
\begin{tabular}{c c c}
\hline\hline
Date & Number of images & Image target \\
\hline
2019-04-11 & 77 & 00360 (Carlova) \\
2019-04-12 & 108 & 00360 (Carlova) \\
2019-04-13 & 77 & 00360 (Carlova) \\
2019-04-26 & 51 & 00145 (Adeona) \\
2019-04-27 & 262 & 00145 (Adeona) \\
2019-04-30 & 124 & 00068 (Leto) \\
2019-05-10 & 337 & 00145 (Adeona) \\
2019-05-11 & 280 & 00145 (Adeona) \\
2019-05-12 & 122 & 00218 (Bianca); 00068 (Leto) \\
2019-05-13 & 21 & 00068 (Leto) \\
2019-05-14 & 80 & 00489 (Comacina) \\
2019-06-03 & 98 & 00031 (Euphrosyne) \\
2019-06-04 & 84 & 00031 (Euphrosyne) \\
2019-06-05 & 66 & 00031 (Euphrosyne) \\
2019-06-06 & 251 & 00218 (Bianca) \\
2019-06-07 & 17 & 00134 (Sophrosyne) \\
\hline
\end{tabular}
\end{table}

Each FITS image measures 1092~$\times$~736 pixels and includes World Coordinate System (WCS) headers for astrometric calibration. Exposure times ranged from 8 to 120 seconds, achieving approximate limiting magnitudes between 19 and 20 under typical observing conditions. Observations were conducted under varying seeing conditions and sky brightness levels.

All 2055 images were calibrated using flat-field and dark frames matching the same exposure times. Subsequently, the images were manually inspected by multiple analysts to determine the presence of linear streaks. Across the dataset, a total of 765 streaks were identified, with measured lengths ranging from 8.5 pixels to 1161.7 pixels, and a mean length of 203.5 pixels.

\begin{figure}[h!]
\centering
\includegraphics[width=\hsize]{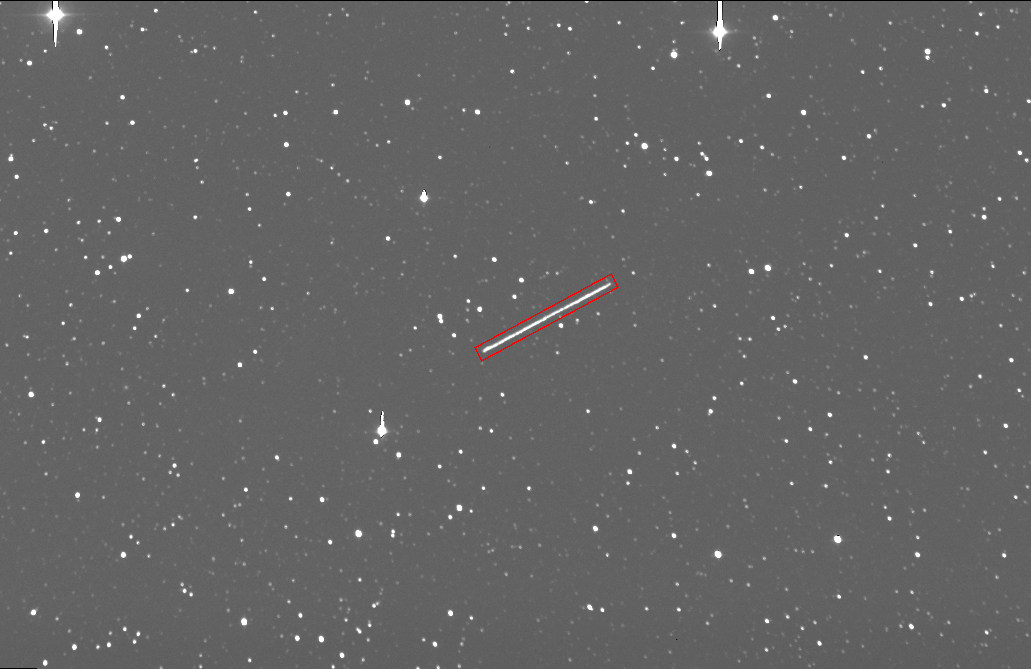}
  \caption{Real image from the dataset obtained during the observation on the night of May 10, 2019, with the La Sagra telescope. An OBB has been overlaid to illustrate the resulting annotation as applied to actual observational data.}
     \label{OBB_gen_dataset}
\end{figure}

The distribution of streak lengths in the real dataset is highly skewed, with a substantially smaller number of long streaks compared to short ones. To enable a controlled stratification of the dataset (described in Section~\ref{subsec:stratification}), the 75th percentile of the streak-length distribution, corresponding to 269.1 pixels, was adopted as the threshold separating short and long streaks. In cases where an image contained multiple streaks, the average streak length within that image was used, since the object of study is the image itself rather than individual streaks. Based on this criterion, the dataset comprises 1523 images without streaks, 412 images containing short streaks, and 120 images containing long streaks. Owing to the resulting imbalance between the short- and long-streak classes, synthetic trails were introduced to increase the representation of long streaks in the dataset. Figure~\ref{OBB_gen_dataset} shows a representative example from the dataset, illustrating an OBB annotation overlaid on a real observational image containing a linear streak.

For each observing night, all images were aligned to a common reference frame in order to streamline the subsequent reduction process. This procedure was applied uniformly across the dataset. As a consequence of small pointing variations and tracking drifts of the telescope, only the region common to all aligned images was preserved, resulting in alignment-induced dead margins in the final images, as discussed in Sect.~\ref{subsec:completeness}.

\subsection{Synthetic data generation}\label{subsec:synthetic_data}

To address the limitations identified in the real FITS dataset, particularly the under-representation of long streaks, synthetic streaks were generated and incorporated into the training set as described below. Synthetic streaks were added to real images using custom Python scripts. For five empirically defined brightness levels, ranging from faint to bright, either full-crossing or partial linear features of random length were injected into the images. A second streak was added in 10\% of the cases to simulate multiple-satellite passages. All synthetic streaks were placed at random locations within the image. A minimum synthetic length of 269\,px—corresponding to the 75th percentile of the real streak-length distribution—was imposed to ensure an adequate representation of long streaks in the augmented dataset. 

In contrast, the angular distribution of the synthetic streaks was designed to resemble the real dataset. Random orientations were drawn from a Gaussian distribution, with 70\% of the simulated lines lying between $-60^\circ$ and $60^\circ$, reflecting the dominant range observed in the real measurements.

To correctly simulate the sensor's effect, the point spread function (PSF) was estimated for each image using PSFEx (PSF Extractor) \citep{Bertin_2013}. The resulting PSF was then convolved, via Fourier transforms, with a one-pixel-wide linear feature to simulate the appearance of streaks on the detector.

A total of 280 additional images containing synthetic streaks were incorporated into the dataset to improve the model performance in identifying long streaks. Among these images, approximately 69\,\% (193) contain synthetic streaks that cross the entire image from side to side, while the remaining 31\,\% (87) contain streaks that remain fully within the field of view.\\

\subsection{Labelling and FITS $\to$ PNG conversion}\label{subsec:labeling}

Having described both the real and synthetic data components, we now detail how the annotated information is prepared for use by the detection model. This subsection describes the manual labelling procedure, the conversion of FITS images into PNG format, and the construction of pixel-based OBBs used for training and evaluation. Annotations were performed manually using the Tycho Tracker software \citep{Parrot_25}.

For each detected streak, a comprehensive set of parameters was meticulously measured and documented, reflecting both astrometric and photometric characteristics as well as geometric details of the streak. The following properties were individually determined for every streak segment: 

\begin{itemize}
    \item Date – Date of observation. (e.g. 2019 04 12.04618).
    \item Observatory – Name of the observatory (e.g. La Sagra - IAA).
    \item MPC\_code – MPC observatory code (e.g. L98).
    \item Telescope – Telescope used for the observation (e.g. TETRA 1 - Reflector f/2 - 0.35m).
    \item Sensor – Detector or camera model (e.g. SBIG ST-10 3 CCD Camera).
    \item Image Name – File name of the image, including date and additional identifiers such as object number, sequence, repetition, chip, and filter (e.g. 20190411\_00360-S003-R001-C001-NoFilt).
    \item Streak Multiplicity – Indicates whether the streak is unique in the frame or if multiple streaks are present (e.g. Single Streak).
    \item Streak Completeness – Indicates whether the streak is complete or incomplete.
    \item Track\_Id – Internal identifier assigned to the streak (e.g. ABC0021).
    \item Central\_RA – Right Ascension of the streak’s centre point (e.g. 16 48 47.16).
    \item Central\_DEC – Declination of the streak’s centre point (e.g. -07 47 35.8).
    \item RA\_Mark1 – Right Ascension of the streak’s starting point (e.g. 16 49 42.20).
    \item DEC\_Mark1 – Declination of the streak’s starting point (e.g. -07 49 37.2).
    \item RA\_Mark2 – Right Ascension of the streak’s end point. (e.g. 16 47 51.82).
    \item DEC\_Mark2 – Declination of the streak’s end point (e.g. -07 45 35.0).
    \item Pixel\_Mark1 – Pixel coordinates (x, y) of the streak’s starting point in the image frame -- for example, (87.00, 670.50).
    \item Pixel\_Mark2 – Pixel coordinates (x, y) of the streak’s end point in the image frame -- for example, (487.50, 709.00).
    \item Central\_Pixel – Pixel coordinates (x, y) of the streak’s centre point -- for example, (286, 689).
    \item Angle – Orientation of the streak with respect to the image Y-axis, measured from 0° to 180° (e.g. 98.5°).
    \item Magnitude – Measured brightness of the streak. In the present implementation, no magnitude is computed. The images were acquired without a physical filter and no photometric transformation was applied. The magnitude field in the MPC line is therefore filled with a placeholder value (e.g. `XX.X R') and is not used in the analysis. The designation `R' is included only as a conventional label, since the unfiltered CCD response tends to be closer to the red part of the spectrum.
    \item MPC Line – Standard MPC-formatted observation record to be reported, containing the object designation, observation date and time in UTC, right ascension (RA) and declination (Dec) of the streak, magnitude, and observatory code. This format allows for uniform reporting of minor planet and small-body observations to the MPC (e.g. ABC0021 C2019 04 12.04618 …).
\end{itemize}

While the S/N is a fundamental parameter for assessing detectability, it was not explicitly logged for this first iteration of our code. Instead, we rely on the magnitude as the primary metric. This choice is justified by the uniformity of the observational data. Given the consistent imaging conditions, parameters and background levels across the dataset, the magnitude serves as an effective proxy for the relative visibility of the streaks. The development of a specific module to calculate and integrate explicit S/N values is ongoing work for future iterations of the pipeline (see Section \ref{subsect:future_work}).

Following the manual labelling process, it was essential to transform the extracted measurements into a format suitable for the training object detection machine learning model. Although YOLO and similar frameworks support a variety of standard image formats (e.g. PNG, JPEG, BMP), they do not natively accept FITS files, nor do they operate on astronomical coordinate systems. Therefore, it was necessary to convert the FITS images to a compatible format and translate the astrometric measurements into pixel-based annotations suitable for model training.

Consequently, a series of pre-processing steps was devised to convert the manually annotated data into structured inputs for the detection pipeline. This included transforming the FITS images into PNG format for compatibility with common computer vision tools and computing precise OBBs around each identified streak. These OBBs ensured that both the spatial extent and the orientation of each streak were accurately preserved in the dataset.

The conversion from FITS to PNG is performed preserving the native orientation of FITS files. This ensures that the visual representation of each PNG matches the original astronomical frame, where the origin is located at the bottom-left and the vertical axis increases upward. Nonetheless, the saved PNG image adheres to the standard raster convention, with the origin at the top-left and the vertical axis increasing downward.

To maintain consistency with this coordinate system, a vertical flip is applied to the $y$-coordinates when loading PNG images for annotation. This adjustment ensures that all geometric computations—such as rotation, scaling, and vertex ordering—are carried out in the reference frame expected by object detection frameworks and image rendering tools.

\medskip
\noindent Special consideration for image sources.  
For real astronomical images, the conversion from FITS to PNG is performed placing the first row of the array at the bottom of the PNG. Consequently, when generating OBBs from FITS-based coordinates (originally referenced with the origin at the top-left), a vertical correction is applied as
\[
y_{\mathrm{corr}} = H_{\mathrm{img}} - y_{\mathrm{original}},
\] 
where \(H_{\mathrm{img}}\) denotes the height of the image in pixels, and \(y_{\mathrm{original}}\) corresponds to the original FITS-based vertical coordinate.
For boosted images, i.e. synthetically generated images introduced for dataset augmentation (hereafter referred to as the synthetic set, also denoted as the boosted set), the pre-processing pipeline directly places the first row at the bottom. Since this vertical inversion is already performed at this stage, the OBB generation step uses
\[
y_{\mathrm{corr}} = y_{\mathrm{original}},
\]
thus avoiding a double flip and ensuring consistent spatial alignment between the image and its annotations.

Throughout the conversion process, the pixel grid remains unaltered, and a ZScale normalisation is applied using the Astropy implementation \citep{2013A&A...558A..33A}, based on the original algorithm from the IRAF data reduction system \citep{Tody_86}. This algorithm enhances the contrast of faint structures by determining the display limits ($z_1, z_2$) through a linear fit to the central pixels of the sorted intensity distribution, thereby excluding extreme outliers. Consequently, all coordinates remain directly comparable between FITS and PNG formats, facilitating accurate streak localisation and annotation.

Bounding boxes for each detected streak are constructed using the pixel coordinates of its two endpoints, denoted as \texttt{Pixel\_Mark1} $(x_1, y_1)$ and \texttt{Pixel\_Mark2} $(x_2, y_2)$. The following procedure defines the OBB associated with each streak.

\begin{figure}[h!]
\centering
\includegraphics[width=\hsize]{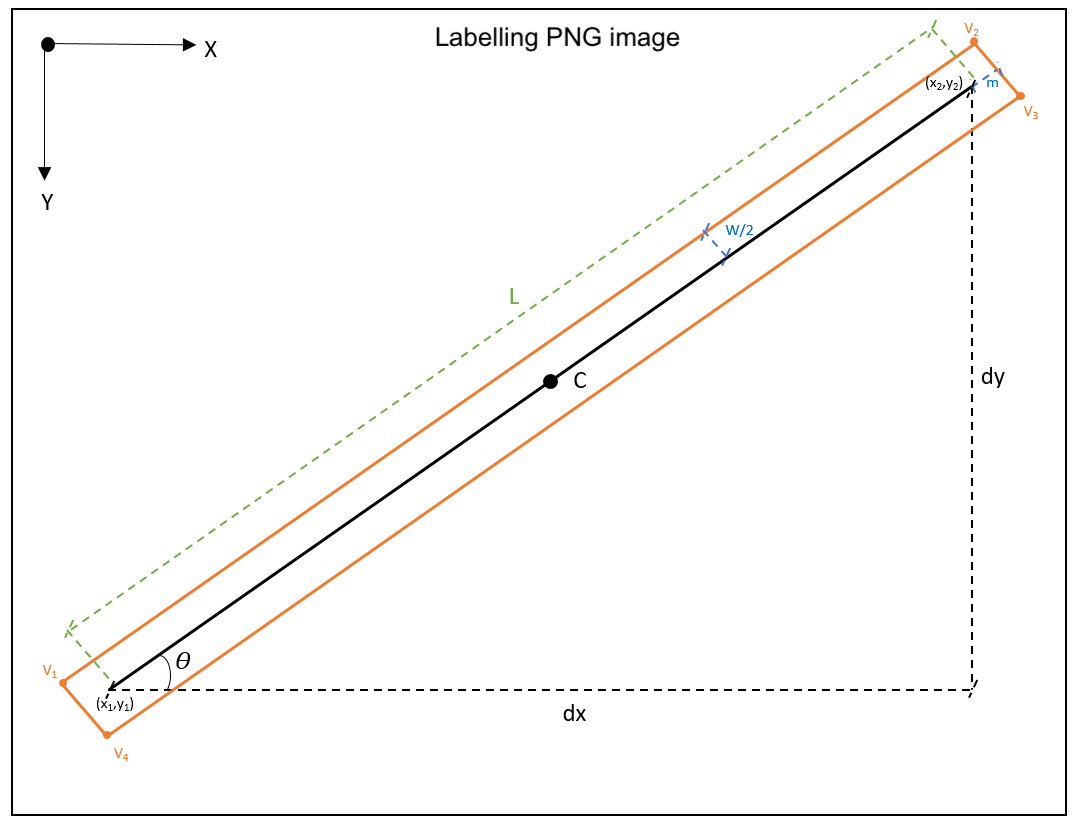}
  \caption{Schematic illustration of the key parameters involved in the bounding box definition.}
     \label{OBB_gen}
\end{figure}

From the two measured streak endpoints, the displacement components $dx$ and $dy$ define the streak length $L$ and its orientation $\theta$ in the image plane. 
The midpoint of the streak defines the OBB centre $C$. 
A fixed margin $m$ is applied along the streak axis to extend the box longitudinally and provide contextual background, while a constant width $W$ defines its transverse extent. 
The four OBB corners $V_k$ are then constructed in a local coordinate system aligned with the streak axis and centred at $C$, and subsequently expressed in the image reference frame. 
When required, the box is clipped to ensure that all vertices remain within the image boundaries, and the final vertex ordering is enforced to match the format expected by YOLO-based detection frameworks in the PNG reference frame. 
A complete mathematical description of this procedure, including clipping and vertex ordering, is provided in Appendix~\ref{app:obb}. 
The geometric parameters involved in this construction are schematically illustrated in Fig.~\ref{OBB_gen}, whereas Fig.~\ref{OBB_gen_dataset} displays a real example from the dataset.

The resulting OBB annotations are thus fully defined in the PNG image reference frame and ready for use in the detection pipeline. These annotated data constitute the basis for the dataset partitioning strategy described in the following subsection.

\subsection{Dataset stratification}\label{subsec:stratification}

This subsection describes the stratification strategy adopted for dataset partitioning and training, ensuring that the training, validation, and test subsets are representative of the same streak-length distribution, both for the initial split of the real images and for the final configuration after synthetic data augmentation.

Before implementing any stratification strategy, we initially trained the model using randomly sampled data. The results obtained from this preliminary setup were acceptable: the model performed as expected in validation and test sets, particularly for streaks that shared characteristics similar to those in the training set. However, this random sampling approach introduced a key limitation, the inability to control the distribution of streak lengths across the training, validation, and test subsets. This posed a risk of bias, especially if long or short streaks were underrepresented in any split.

To support the stratified sampling strategy used later in the training pipeline, we first analysed the distribution of individual streak lengths. As previously described in Section~\ref{subsec:fits}, the dataset contains a total of 765 real streaks, with lengths covering a broad dynamic range. For convenience, Table~\ref{table:percentiles} summarises the main descriptive statistics and percentiles of this distribution.

\begin{figure}[h!]
\centering
\includegraphics[width=\hsize]{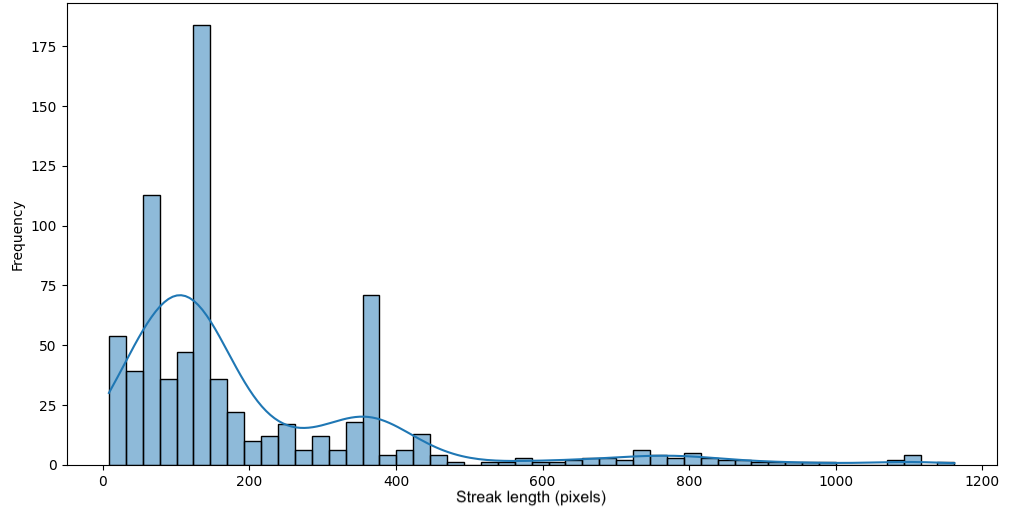}
  \caption{Histogram of the individual streak lengths across the dataset.}
     \label{fig:length_histogram}
\end{figure}

The 75th percentile (269.1 pixels) was adopted as the operational threshold separating short and long streaks, a choice motivated by the clear skew of the distribution (Figure~\ref{fig:length_histogram}). This threshold provides a practical balance between representativeness and class separability, enabling a controlled stratification of the dataset for training, validation, and test splits.

\begin{table}[H]
\caption{Length percentiles for all streaks in the dataset.}
\label{table:percentiles}
\centering
\begin{tabular}{c c}
\hline\hline
Metric & Value \\
\hline
25th percentile (P25)     & 75.2 px   \\
50th percentile (P50)     & 127.4 px  \\
75th percentile (P75)     & 269.1 px  \\
85th percentile (P85)     & 362.4 px  \\
90th percentile (P90)     & 405.7 px  \\
95th percentile (P95)     & 716.6 px  \\
\hline
\end{tabular}
\end{table}

Such stratification is particularly relevant to mitigate potential overfitting to specific streak lengths and to enhance the generalisation capability of the model under varying observational conditions. The full distribution of streak lengths in the real dataset is illustrated in Figure~\ref{fig:length_histogram}, where a clear skew towards shorter streaks is visible, with a long tail representing a minority of very long detections.

After selecting the 75th percentile as the threshold for streak length, we proceeded to classify the images in the dataset into three categories based on the average length of the streaks they contain. Specifically, we identified 412 images classified as short-streak images, 120 as long-streak images, and 1523 images without streaks. For images containing both short and long streaks, we computed the mean streak length and applied the threshold criterion accordingly: if the average streak length exceeded 269.1 pixels, the image was classified as containing long streaks; otherwise, it was considered a short-streak image. It is important to note that the detection algorithm is able to detect multiple streaks in a single image; the use of the mean streak length per image is only introduced here as a convenient summary statistic for the stratification procedure and does not reflect any limitation on the number of streaks that can be detected.

To ensure a balanced representation of the three image-level classes across all dataset partitions, we applied a stratified splitting strategy using the image classification described above. The complete dataset was divided into training, validation and test subsets with proportions of 70\%, 20\%, and 10\%, respectively.

The real image dataset split consists of 1438 images in the training set, 411 images in the validation set, and 206 images in the test set. Each subset preserves the original class distribution of short-streak, long-streak, and no-streak images, as ensured by the stratified splitting strategy. This balanced distribution across splits is essential to prevent training bias and to obtain reliable and consistent evaluation metrics during model validation and testing, as summarised in Table~\ref{tab:split_distribution}.

\begin{table}[H]
\centering
\caption{Number of images and class distribution in each subset of the dataset of real images.}
\label{tab:split_distribution}
\begin{tabular}{lcc}
\hline\hline
Subset & Class & Count (\%) \\
\hline
\multirow{3}{*}{Training}   & Long       & 84 (5.8\%)   \\
                            & Short      & 288 (20.0\%) \\
                            & No-streak  & 1066 (74.2\%) \\
\hline
\multirow{3}{*}{Validation} & Long       & 24 (5.8\%)   \\
                            & Short      & 83 (20.2\%)  \\
                            & No-streak  & 304 (74.0\%) \\
\hline
\multirow{3}{*}{Test}       & Long       & 12 (5.8\%)   \\
                            & Short      & 41 (19.9\%)  \\
                            & No-streak  & 153 (74.3\%) \\
\hline
\end{tabular}
\tablefoot{Percentages are relative to the corresponding subset.}
\end{table}

Although the stratification procedure described above ensured that each subset preserved the original proportions of short-streak, long-streak, and no-streak images contained in the real dataset, the resulting distribution remained imbalanced with respect to the relative frequency of long streak versus short streak images. Specifically, short-streak images account for roughly $\sim$20\% of each subset, whereas long-streak images constitute only $\sim$6\%. This imbalance may hinder the model’s ability to learn robust features for underrepresented streak types, particularly long streaks and thus potentially limit its generalisation capability.

To mitigate this limitation, synthetic data generation was applied as described in Section~\ref{subsec:synthetic_data}. The resulting augmented dataset (real plus synthetic) achieved a more balanced representation of short-streak and long-streak images, while slightly reducing the proportion of no-streak images. The final distribution of the combined dataset is summarised in Table~\ref{tab:split_distribution_augmented}.

\begin{table}[H]
\centering
\caption{Number of images and class distribution in each subset of the final (real plus synthetic) dataset.}
\label{tab:split_distribution_augmented}
\begin{tabular}{lcc}
\hline\hline
Subset & Class & Count (\%) \\
\hline
\multirow{3}{*}{Training}   & Long       & 262 (16.0\%)   \\
                            & Short      & 306 (18.7\%)   \\
                            & No-streak  & 1066 (65.2\%)  \\
\hline
\multirow{3}{*}{Validation} & Long       & 75 (16.1\%)    \\
                            & Short      & 88 (18.8\%)    \\
                            & No-streak  & 304 (65.1\%)   \\
\hline
\multirow{3}{*}{Test}       & Long       & 37 (15.8\%)    \\
                            & Short      & 44 (18.8\%)    \\
                            & No-streak  & 153 (65.4\%)   \\
\hline
\end{tabular}
\tablefoot{The final dataset includes both real and synthetic images. Percentages are relative to the corresponding subset}
\end{table}

This final stratified and augmented dataset constitutes the input used for training, validation, and testing of the detection model, whose performance is analysed in the following section.

\section{Results and integration into the detection database}

The model's performance was first evaluated under controlled conditions using the held-out test set, as described in Section~\ref{subsec:confusion_matrix}. This evaluation, based on standard metrics such as precision, recall, and mean Average Precision (mAP50), provided a quantitative assessment of detection accuracy, complemented by qualitative visual inspection to verify practical effectiveness in detecting streaks under astronomical conditions.

Beyond the controlled evaluation on the test set, the model was applied to real observational data, where additional artefacts and observational effects (e.g. diffraction spikes or saturated stars) required dedicated post-processing and database-driven filtering.

As a preliminary step, inference was run on a set of 273 images using the trained model with an input resolution of 640~px, a confidence threshold\footnote{Minimum confidence required for a predicted object to be retained; lower values increase sensitivity but may introduce false positives.} of 0.25, and an IoU threshold of 0.45 for box matching. The complete workflow comprises: (i) bright-star filtering and extraction of temporal metadata (Section~\ref{subsec:filter_spikes}), (ii) photometric extension of OBBs to recover the full streak length (Section~\ref{subsec:obb_elongation}), (iii) estimation of streak endpoints from the refined corner sets (Section~\ref{subsec:corner_clustering}), (iv) geometric extrapolation and inter-frame association of detections belonging to the same physical object (Section~\ref{subsec:interframe_association}), and (v) record standardisation and assessment of streak completeness (Section~\ref{subsec:completeness}).

\subsection{Performance metrics}
\label{subsec:confusion_matrix}

To assess the performance of the trained model, we performed both a quantitative evaluation on the test set and a qualitative visual inspection of selected predictions. This dual approach allowed us to not only measure the model's accuracy using standard metrics but also verify its practical effectiveness in detecting streaks under real astronomical conditions.

\begin{figure}[h!]
\centering
\includegraphics[width=\hsize]{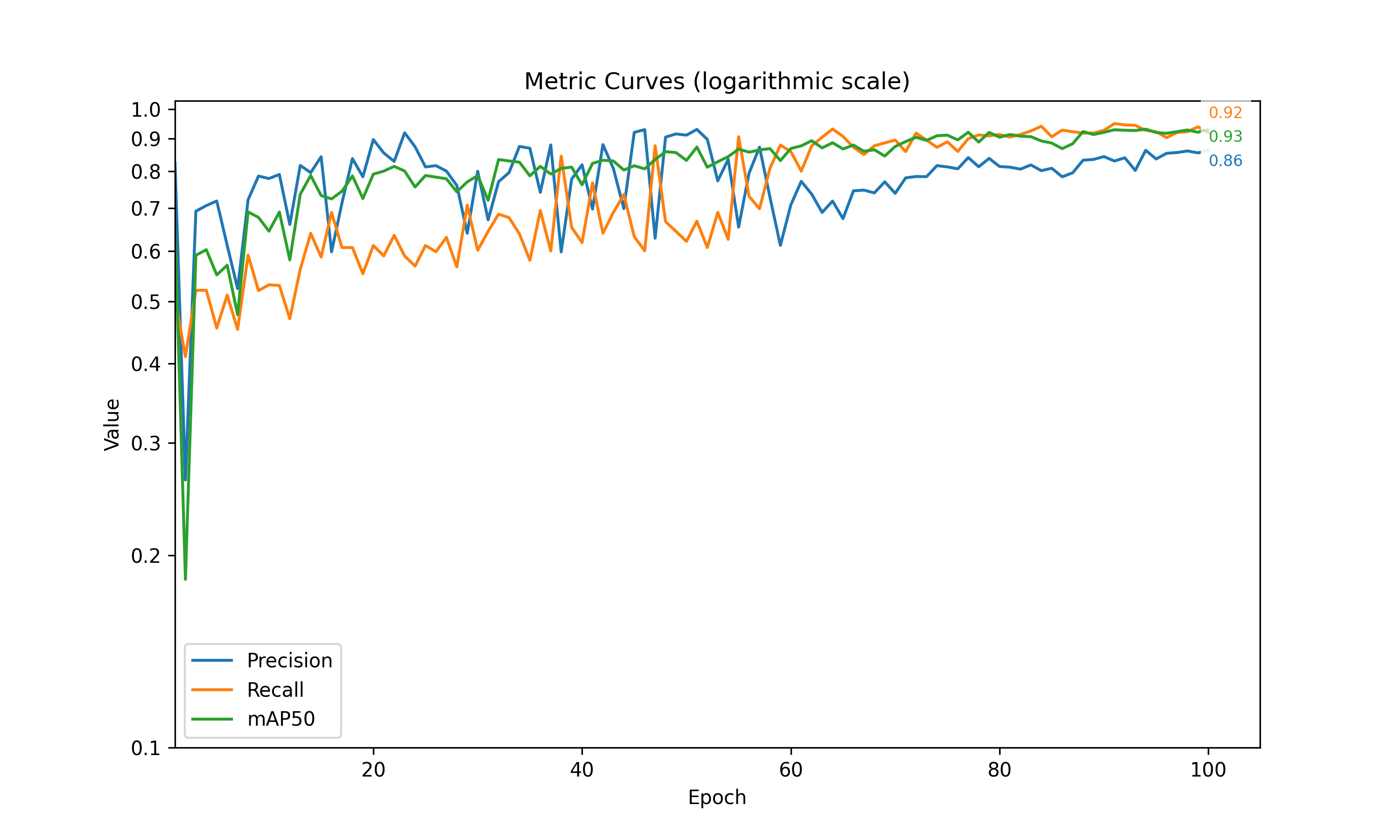}
    \caption{Evolution of performance metrics—precision, recall, and mAP50—throughout the 100 training epochs. The values are plotted on a logarithmic scale to emphasise performance dynamics over time.}
     \label{fig:metrics_log}
\end{figure}

Model performance was evaluated using standard metrics such as precision\footnote{Precision is defined as TP $/$ (TP + FP), where TP denotes true positives and FP denotes false positives. It reflects the proportion of correct positive predictions among all positive predictions made by the model.}, recall\footnote{Recall is defined as TP $/$ (TP + FN), where TP denotes true positives and FN denotes false negatives. It reflects the proportion of actual positives that were correctly identified by the model.}, and mean Average Precision at Intersection over Union (IoU\footnote{The Intersection over Union (IoU) measures the overlap between the predicted bounding box and the ground truth box. It is defined as the area of their intersection divided by the area of their union, and it is commonly used to determine whether a detection is considered correct.}) 0.5 (mAP50) \citep{Goodfellow_16, Everingham_10}. Figure~\ref{fig:metrics_log} shows the evolution of these three metrics across the training epochs.

\begin{figure}[h!]
\centering
\includegraphics[width=\hsize]{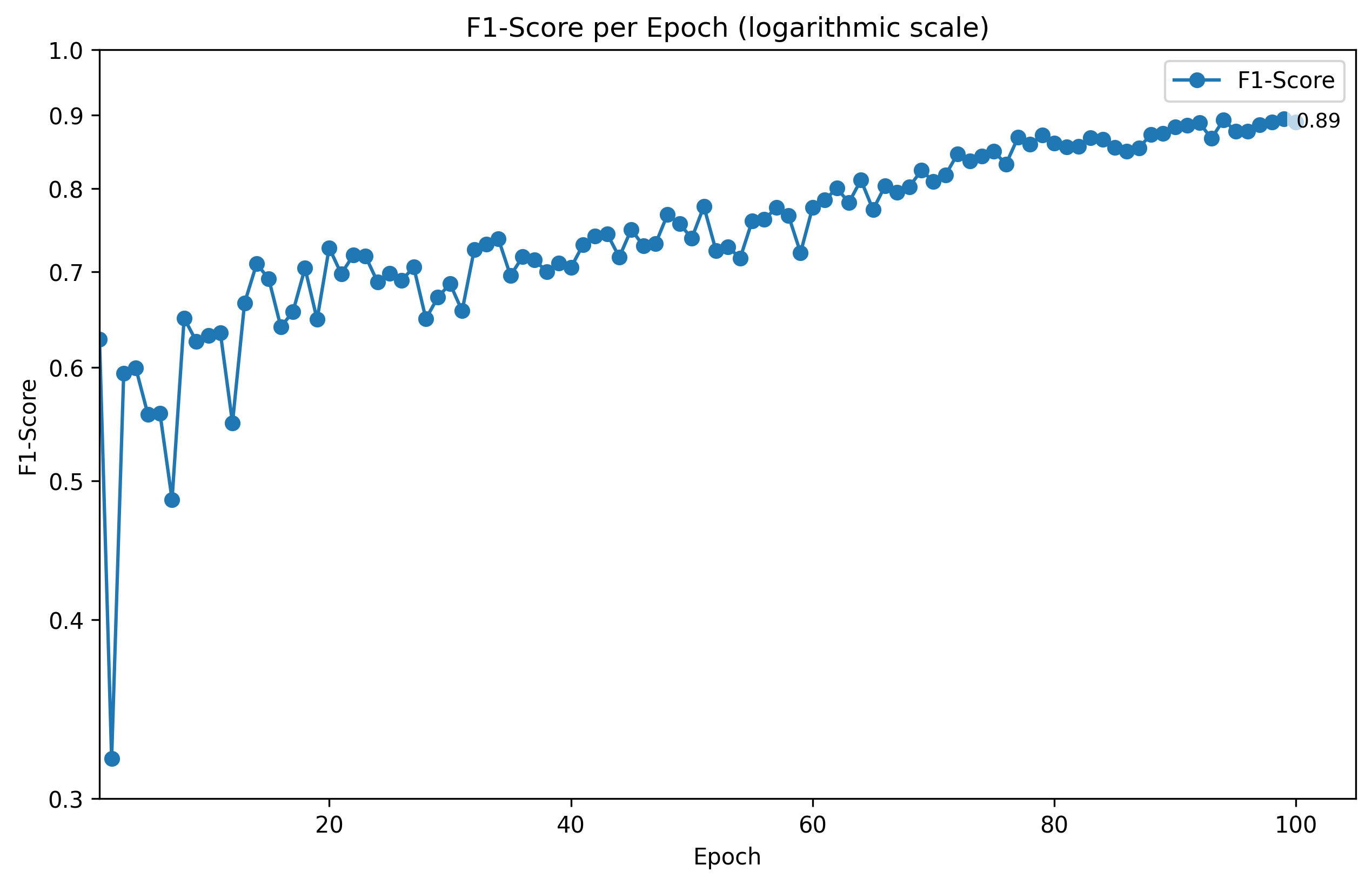}
    \caption{Evolution of the F1-score per training epoch using a logarithmic scale on the vertical axis. The curve highlights the overall classification performance of the model across epochs. }
     \label{fig:F1}
\end{figure}

In addition to standard performance metrics, the F1-score\footnote{The F1-score is defined as $2 \cdot \frac{\text{Precision} \cdot \text{Recall}}{\text{Precision} + \text{Recall}}$. It represents the harmonic mean of precision and recall, and penalises models that disproportionately favour one over the other.} \citep{Goodfellow_16} was computed to provide a balanced measure of classification performance. This metric combines both precision and recall into a single value, allowing for a more comprehensive evaluation of detection quality. Its progression across training epochs is illustrated in Figure~\ref{fig:F1}.

In order to assess the classification behaviour of the trained model, a
normalised confusion matrix was computed on the test set using a fixed IoU
threshold of 0.8 to determine detection matches. The test set consists of held-out images strictly excluded from the training and validation process, serving as an unbiased benchmark to evaluate the model's generalisation capabilities.
This matrix summarises how the
predictions are distributed across the two binary classes (streak / no streak).
The left column is expressed in number of individual trails, while the right
column refers to number of images, because background frames do not contain
bounding-box annotations and must therefore be evaluated per image.

\begin{figure}[h!]
\centering
\includegraphics[width=\hsize]{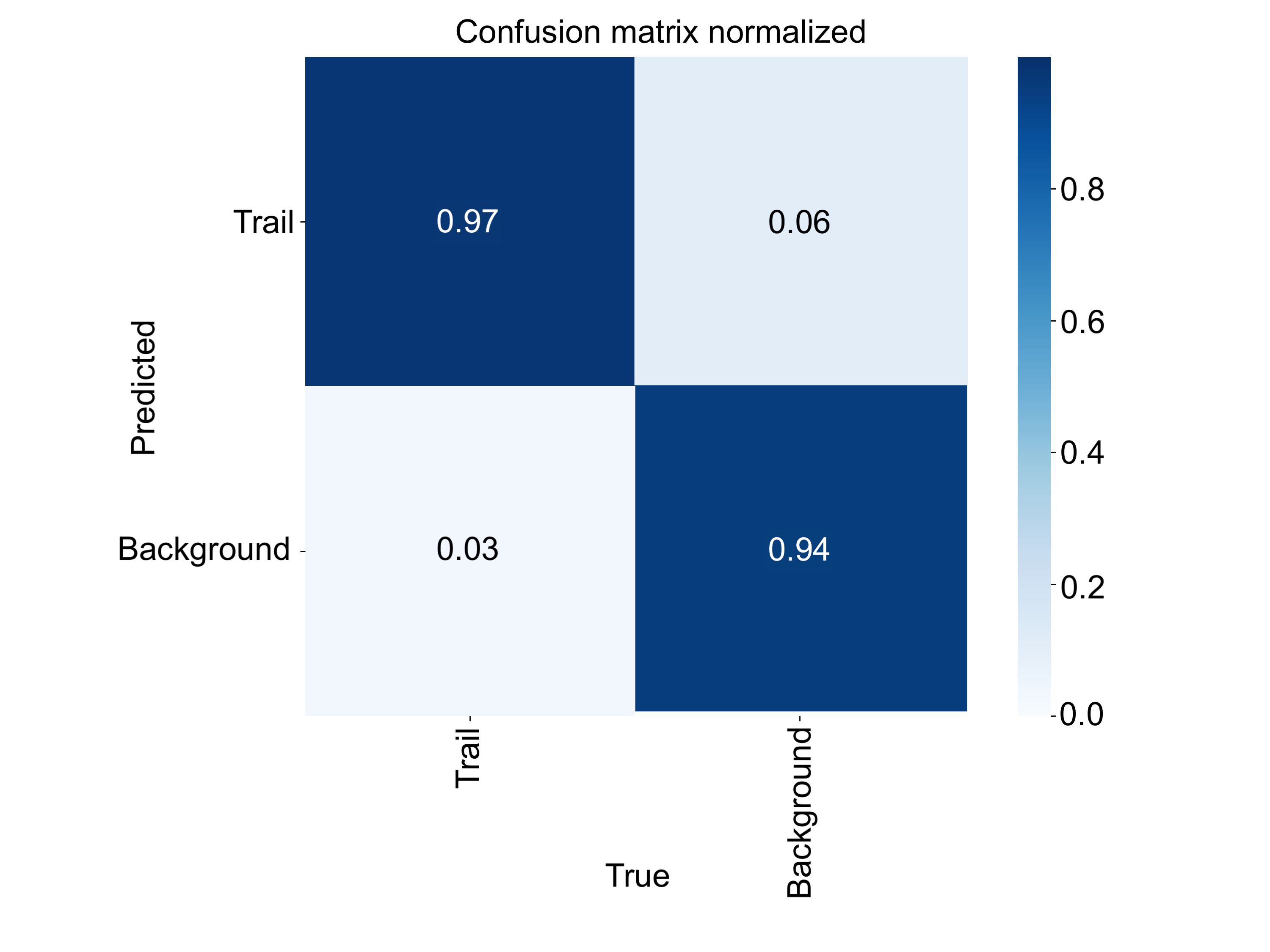}
\caption{Normalised confusion matrix evaluated on the test set using an IoU
threshold of 0.8. The left column is expressed in trails, the right column in
images, as background frames lack bounding-box annotations and were evaluated per
image.}
\label{fig:confusion_matrix}
\end{figure}

This evaluation uses a hybrid definition reflecting the nature of the available
labels. On the trail side (left column), each physical streak is counted
as detected if at least one predicted OBB overlaps the ground-truth annotation
with IoU $\geq 0.8$. Under this criterion, the model correctly identifies 107 of
110 real streaks (TP = 107, FN = 3).

On the image side (right column), false positives are counted per image:
a background frame is considered a false positive if the model produces a streak
in an image where no real streak is present. All false-positive detections were verified by visual inspection, and in every case the spurious prediction appears in a different background frame. This one-to-one relation between false positives and images enables a consistent estimate of the number of true-negative frames in the test set.

Taking this into account, and in order to provide a fully populated mixed-unit
confusion matrix for reference, the number of true-negative images is obtained
directly from the composition of the test set. The background subset contains
153 images, and every false positive appears in a different background frame.
Therefore, subtracting the six false-positive images yields a total of
147 true-negative images (Figure ~\ref{fig:confusion_matrix}). Using the strict IoU-based definition as the benchmark, the model achieves a precision of 94\,\% and a recall of 97\,\% on the test set. These results demonstrate that the detection model achieves high accuracy under controlled evaluation conditions.

\subsection{Pre-processing: Stellar filtering and temporal metadata extraction}
\label{subsec:filter_spikes}

Under realistic observing conditions, and particularly in frames containing very bright stars, the baseline detection model occasionally misclassified horizontal stellar diffraction spikes as streaks, whereas vertical spikes were largely unaffected by this behaviour. An example is shown in Fig.~\ref{fig:obb_elongation_example}, where a high-confidence prediction would be erroneously assigned to a horizontal spike emerging from a saturated star. To suppress these false positives, a catalogue-driven filtering stage was implemented.

A preliminary survey of the images was used to determine the magnitude threshold above which saturated stars produce detectable diffraction spikes. Based on this, a catalogue-driven filtering procedure was implemented: for each image zone, a cone search against the Gaia~DR3 catalogue \citep{Gaia_23} is performed via ADQL queries, retrieving all stars brighter than the adopted threshold within a padded region. The resulting stellar catalogue is then cross-matched with the candidate detections using the \texttt{search\_around\_sky} function from \texttt{Astropy} \citep{2013A&A...558A..33A}. Any predicted box located within a predefined angular distance of such stars is flagged as contaminated and removed from further processing. Catalogue queries are parallelised across zones and the filtering is performed in a fully vectorised manner, ensuring that nearly all spike-induced false positives are eliminated while genuine streaks are preserved.

After applying the bright-star filter, the number of spurious detections is substantially reduced. Out of the 66 spike-induced false positives initially present in the inference set (129 detections in total), the filter removes 51, leaving only 15 residual cases. This corresponds to a filter efficiency of 77\,\%, which is sufficient to prevent these artefacts from significantly impacting the subsequent processing stages.

In addition to suppressing spike-induced false positives, this pre-processing stage also extracts the temporal metadata required for subsequent inter-frame analysis. FITS headers are parsed for all observing sequences to extract instrumental and
temporal metadata. From these headers, three products are constructed: the
image dimensions, the exposure time, and the table of inter-frame intervals computed from
successive entries.\footnote{These intervals represent the
elapsed time between the start of consecutive exposures and include both the
exposure time and the inter-exposure idle time.} These intervals provide the
temporal baseline required for the endpoint-propagation analysis discussed in
Sect.~\ref{subsec:interframe_association}. 

Spike-induced false positives are largely suppressed during the filtering stage. The remaining detections are then geometrically refined to improve the representation of genuine streaks before inter-frame association.

\subsection{OBB longitudinal extension via photometric pre-analysis}
\label{subsec:obb_elongation}

With the spurious detections removed, the next step focuses on refining the geometry of the remaining streak candidates prior to inter-frame association. The OBBs returned by the model frequently underestimate the true extent of linear streaks, as YOLO-based regressors are not optimised for objects that span a significant fraction of the field of view. To address this limitation, an additional post-processing stage performs a photometric pre-analysis along the OBB’s major axis and extends the box longitudinally while ensuring geometric consistency.

For each selected detection, the corresponding PNG frame is loaded and transformed into a photometrically enhanced image via greyscale conversion, Contrast Limited Adaptive Histogram Equalization (CLAHE), Gaussian smoothing, and normalisation. This pre-processing increases the contrast of faint streak wings and stabilises the subsequent background estimation.

From the original OBB, the main direction of the trail, its transverse direction, its width, and its two endpoints are recovered. A one-dimensional flux profile \(I(s)\) is then sampled along the axis using bilinear interpolation. Instead of measuring a single line, the profile is computed as the mean over a small bundle of parallel strips spanning the transverse width of the streak, which improves robustness against pixel noise, PSF asymmetries, and local defects.

The background level \(I_{\mathrm{bg}}\) and the noise \(\sigma\) are estimated from two sidebands parallel to the streak, displaced outwards by a fixed margin and sampled across a transverse band to obtain a representative distribution of background intensities. The elongation proceeds while the measured flux remains above a dynamic threshold,
\[
I(s) > I_{\mathrm{thr}} = I_{\mathrm{bg}} + k\,\sigma ,
\]
where the factor \(k\) is adapted to the local signal-to-noise ratio measured in short inner segments located immediately inside each end of the streak. These segments are evaluated using the same multi-strip approach in order to stabilise the S/N estimate. Low-S/N extremities yield higher values of \(k\), preventing over-extension driven by isolated fluctuations, while allowing genuinely faint parts of the trail to be recovered.

The endpoints are then advanced iteratively in the main direction, one pixel at a time. The propagation continues as long as the intensity remains above threshold and several safeguards are satisfied: a minimum number of consecutive sub-threshold samples, a relative cut based on a fraction of the profile’s peak intensity, and a maximum elongation defined as a multiple of the original box length. A final post-check suppresses noise-driven excursions, and elongations that do not exceed a minimal geometric gain are discarded.

\begin{figure}[h!]
\centering
\includegraphics[width=\hsize]{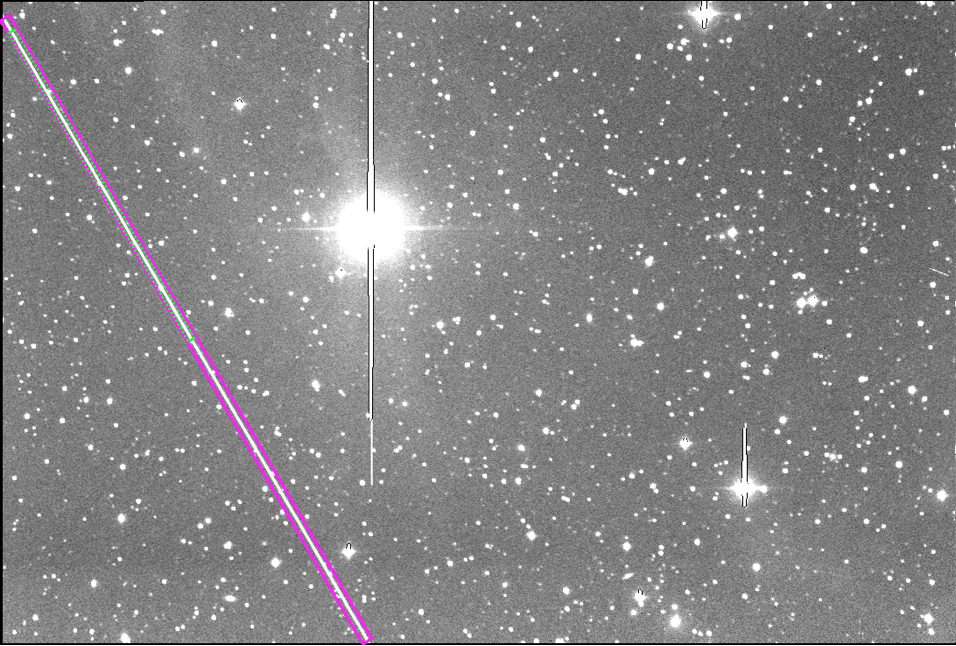}
\caption{Example of the photometric elongation procedure applied to a single detection. The original OBB returned by the model is shown in green, while the photometrically extended OBB is shown in magenta.} 
\label{fig:obb_elongation_example}
\end{figure}

The resulting extended axis defines a new OBB. A deep copy of the detection dictionary is updated accordingly: only the geometric fields of each entry are modified, and a flag is stored indicating whether elongation occurred. An illustrative example of this procedure is shown in Fig.~\ref{fig:obb_elongation_example}, where the original OBB (green) and the photometrically extended OBB (magenta) are displayed for a representative detection. The updated dictionary is then saved and propagated to the subsequent stages of the analysis pipeline.

Following the longitudinal extension of the OBBs, the streak endpoints are determined. These endpoints are used to define the streak centre and to enable subsequent temporal propagation and inter-frame association.

It should be emphasised that this pre-analysis is used exclusively for geometric refinement of the OBBs and for determining reliable streak endpoints. No calibrated photometry of the streaks is performed in this work, and the Gaia catalogue is employed solely for astrometric purposes. Consequently, no total calibrated magnitudes are derived from the streak flux measurements. A dedicated photometric analysis is planned for future developments of StreakMind (see Sect.~\ref{subsect:future_work}).

\subsection{Endpoint estimation from corner clusters}
\label{subsec:corner_clustering}

The model outputs OBBs defined by four corner points. These four vertices, updated after the photometry process described in Sect.~\ref{subsec:obb_elongation}, are partitioned into two groups using agglomerative hierarchical clustering with Ward linkage, where the grouping is based on minimising the increase in within-cluster variance at each merge step \citep{Ward1963}. In this constrained case, the procedure produces exactly two clusters, each containing the pair of adjacent vertices that define one extremity of the streak. The centroid of each cluster provides a stable estimate of the corresponding endpoint. The streak centre is then defined as the midpoint between the two estimated endpoints. These points serve as reference markers for subsequent geometric and temporal analyses. In parallel, the celestial coordinates (RA/Dec) associated with the four corners are averaged within each cluster to derive endpoint astrometry, which is later converted into sexagesimal notation for database integration.

Once the streak endpoints are defined, they provide the geometric reference required to extrapolate the streak position across consecutive frames. The streak centre is retained as a compact descriptor for astrometric and database purposes.

\subsection{Geometric extrapolation and inter-frame association}
\label{subsec:interframe_association}

The purpose of the inter-frame association stage is to determine whether streaks detected in consecutive images correspond to the same physical object. The procedure accounts for the possibility that a streak may continue in either direction along its original axis in subsequent frames. To this end, two extrapolation hypotheses are considered: one projecting the markers forwards from the trailing endpoint, and another projecting them backwards from the leading endpoint. The process begins with a preliminary boundary check: if both predicted markers fall outside the valid image domain under both extrapolations, the track is regarded as unique and no further association is pursued. In such cases, the corresponding object is interpreted as producing only that single streak within the observed passage. 

If this termination condition is not met, the subsequent step predicts the likely marker positions in following frames. For a detection with pixel markers $\mathbf{m}_1,\mathbf{m}_2 \in \mathbb{R}^2$, the segment is extrapolated to the image bounds to generate a dense set of candidate pixels along the inferred trail direction. Let $\Delta \mathbf{m}=\mathbf{m}_2-\mathbf{m}_1$, and $t_{\exp}$ be the exposure time. A pixel-velocity estimate is defined as
\[
\mathbf{v} \;=\; \frac{\Delta \mathbf{m}}{t_{\exp}} \quad [\text{px s}^{-1}].
\]

The temporal offset to a future frame ($k$) is obtained by accumulating the inter-frame intervals described in Section~\ref{subsec:filter_spikes}. Each interval represents the elapsed time between the start of two consecutive exposures, comprising both the exposure duration of the earlier frame and the subsequent idle gap. To ensure that the propagation is anchored at the end of the initial exposure rather than its beginning, we subtract the contribution of the first exposure from the cumulative sum. The resulting offset $\Delta t_k$ therefore corresponds to the effective time elapsed between the completion of the reference exposure and the start of frame $k$. This value provides the temporal baseline for projecting the streak forwards in time. To account for potential orientation inversions, we consider both forwards and backwards hypotheses:

\[
\mathbf{m}_1^{(k)}=\mathbf{m}_2 + \mathbf{v}\,\Delta t_k,\quad
\mathbf{m}_2^{(k)}=\mathbf{m}_1^{(k)}+\Delta \mathbf{m},
\]
\[
\tilde{\mathbf{m}}_1^{(k)}=\mathbf{m}_1 - \mathbf{v}\,\Delta t_k,\quad
\tilde{\mathbf{m}}_2^{(k)}=\tilde{\mathbf{m}}_1^{(k)}-\Delta \mathbf{m}.
\]

\noindent \paragraph{Notation.} 
Superscript notation $^{(k)}$ denotes predicted marker positions in frame $k$, 
while tildes ($\tilde{\mathbf{m}}$) indicate the alternative backwards hypothesis 
accounting for possible inversions in the streak orientation. 
In this context, $\mathbf{m}_1$ and $\mathbf{m}_2$ are preserved as symbolic labels for the two streak endpoints, even though their physical roles as entry or exit points may swap under the backwards hypothesis. This convention ensures algebraic consistency in the propagation formulas, while accommodating both orientation scenarios.

For each candidate detection in frame $k$, the orientation angle is first compared to that of the reference streak. Consistency is required within an angular tolerance of $\Delta\theta=6^\circ$. Only candidates passing this angular check proceed to the positional verification, where both of their markers must lie within a pixel tolerance of $\tau=5$~px from the extrapolated segment. A match is established only if both conditions are satisfied simultaneously, in which case the detection is associated with the reference streak and the track is extended across frames. 

If no candidates meet these criteria, the missed-frame counter\footnote{The missed-frame counter allows the association to tolerate up to two consecutive frames without a valid match. This prevents spurious breaks in the track linkage due to transient issues such as clouds, poor seeing, or temporary image degradation, ensuring that detections belonging to the same physical object are not mistakenly split into separate tracks.} is incremented and the procedure advances to the next image. In this work, a threshold of \texttt{max\_miss} = 2 was adopted, meaning that a streak can remain unmatched for tolerating up to two consecutive frames without a valid match; the track is terminated only if no match is found in the third consecutive frame. If both the forwards and backwards extrapolation hypotheses fall entirely outside the valid image boundaries, the association is terminated without further evaluation, as the object is no longer observable within the field of view.

It is important to note that the association procedure is performed independently for each detected streak: every streak is individually propagated across subsequent frames using its own geometric and temporal predictions. This design ensures that simultaneous objects are correctly disentangled, with each one preserving its unique track identifier. For instance, if object A is first detected in frame $I_1$ and object B in frame $I_2$, both will be independently tracked and correctly re-associated in frame $I_3$, where the two streaks coexist. Even in cases where object A is temporarily absent from $I_2$, the missed-frame tolerance ensures that its trajectory is consistently re-linked once it reappears. In this way, the procedure prevents cross-contamination between tracks and guarantees that each multi-frame track corresponds to a unique moving object. The outcome of the association stage is subsequently transformed into standardised records and subjected to a completeness assessment before being incorporated into the detection database.

\subsection{Record standardisation and completeness assessment}
\label{subsec:completeness}

A streak is labelled `complete' if both markers lie strictly inside an inner margin of width $\tau_{\text{edge}}$ from the four image borders, and `incomplete' otherwise. Certain images exhibit alignment-induced dead margins, sometimes exceeding 80~px, due to telescope motion over the course of the night. To mitigate misclassification, a threshold of $\tau_{\text{edge}}=40$~px was adopted: values that are too small would mark incomplete streaks near these margins as complete, whereas overly large values would incorrectly flag truly complete streaks as incomplete. The threshold was empirically selected to balance both considerations across observing nights. Formally, with image bounds $[0,W_{\text{img}})\times[0,H_{\text{img}})$ and $\tau_{\text{edge}}=40$~px,
\[
\tau_{\text{edge}} < x_{m_i} < W_{\text{img}}-\tau_{\text{edge}}, \quad
\tau_{\text{edge}} < y_{m_i} < H_{\text{img}}-\tau_{\text{edge}}, \quad i\in\{1,2\}.
\]

This completeness status is incorporated into the normalised record together with the astrometric solution. Following the inter-frame association procedure described in Section~\ref{subsec:interframe_association}, unique track identifiers (\texttt{Track\_ID}) are assigned sequentially to detections and propagated to associated matches. In addition to per-detection parameters, constant metadata are uniformly attached to all detections belonging to a given observing set, including the observatory name, MPC code, telescope, sensor, field-of-view, observer, and analyst. Central and marker astrometry are stored in sexagesimal format, while pixel coordinates are rounded to two decimals for reproducibility. Image names are normalised to include the observing date (UTC) derived from \texttt{DATE-OBS}.

On this basis, an MPC-style observation line is synthesised using the final \texttt{Track\_ID}, MPC-formatted date, astrometry and site code. The resulting standardised record is appended to the SQL-ready dictionary, ensuring that all detections can be directly ingested into the detection database.

\section{Satellite cross-identification and database export}
\label{subsec:sat_xid_export}

Having established the geometric characterisation, temporal association, and standardised representation of the detected streaks, the final step of the pipeline addresses their physical interpretation. At this stage, each multi-frame detection is evaluated to determine its most likely origin and to quantify the confidence of its identification through cross-identification with external ephemerides and subsequent confidence scoring, prior to database ingestion. The procedure is organised into the following stages:

    \paragraph{Cross-identification against external ephemerides.}

    Each observation is converted into an MPC-style astrometric line (Section~\ref{subsec:completeness}) and submitted in batch to the public Sat\_ID service of Project Pluto \citep{ProjectPluto2025}. The query returns a list of candidate satellites together with their angular offsets from the reported position. A conservative search radius of $1^\circ$ is adopted in all submissions, ensuring that only plausible matches are returned.
    
    Two independent runs are executed: (i) using exclusively the set of complete streaks and (ii) using all the streaks.
    This dual approach mitigates the effect of incomplete streaks, whose truncated morphology can displace the centroid used in the ephemeris query, occasionally suppressing otherwise valid identifications. The results of both runs are merged (after filtering explained at the beginning of the next subsection) by retaining, for each MLD identifier and each satellite code, only the occurrence with the smallest angular offset.\\

    \paragraph{Offset filtering and Gaussian-based confidence scoring.}
    
    Before computing the confidence scores, candidates whose offsets are unreasonably
    large compared with the minimum offset are discarded. A candidate is removed if its offset is more than five times the minimum, or if it exceeds the minimum by more than \(0.4^\circ\). The surviving offsets are then evaluated through a two–component confidence model governed by two Gaussian scales: a relative scale
    \(\sigma_{\mathrm{rel}} = 0.2^\circ\) that measures how quickly confidence decreases for offsets deviating from the minimum, and an absolute scale
    \(\sigma_{\mathrm{abs}} = 0.3^\circ\) that penalises cases in which the minimum offset itself is large.
    
    We let the filtered offsets, sorted in ascending order, be
    \[
    o_0 \le o_1 \le o_2 \le \dots,
    \]
    with \(o_{\min} = o_0\). The first component of the model assigns each offset a 
    relative-distance weight,
    \[
    w_i = \exp\!\left( -\frac{(o_i - o_{\min})^2}{2\sigma_{\mathrm{rel}}^2} \right),
    \]
    which is normalised to
    \[
    p_i = \frac{w_i}{\sum_j w_j}
    \]
    so that candidates close to the minimum dominate the probability distribution.
    
     The second component penalises solutions where the minimum offset itself is large,
    \[
    \mathrm{abs\_factor} =
    \exp\!\left( -\frac{o_{\min}^2}{2\sigma_{\mathrm{abs}}^2} \right).
    \]
    
     The confidence of each candidate is then
    \[
    \mathrm{Conf\_Score\_cand}_i = p_i \cdot \mathrm{abs\_factor},
    \]
    and the global confidence score is defined as
    \[
    \mathrm{Conf\_Score} = p_0 \cdot \mathrm{abs\_factor},
    \]
    where \(p_0\) is the normalised weight associated with the candidate of minimum
    offset (i.e. the first element in the ordered set).
    All confidence values lie in \([0,1]\).\\

    \paragraph{Pass-separation safeguard.}

    A secondary consistency step is applied after the confidence scoring to prevent
    overwriting the confidence values of detections belonging to different passes
    of the same satellite. For each satellite code, detections are examined within
    a small window of neighbouring MLD identifiers. When multiple candidates for the
    same satellite appear inside this window, the global confidence (\texttt{Conf\_Score}) assigned to each
    entry is set to the maximum of the individual per-candidate scores
    (\texttt{Conf\_Score\_cand}) found in that neighbourhood. This ensures that
    detections from the same pass retain coherent confidence values, while
    independent passes remain isolated and are not affected by one another. The result of this process is a unified table where each MLD entry is
    associated with the angular offset to the candidate satellite, the
    candidate satellite code, the per-candidate confidence score, and the local
    global confidence.\\

    \paragraph{Record enrichment and database export.}
    
    The consolidated identification results (offset, satellite code,
    \texttt{Conf\_Score}, and \texttt{Conf\_Score\_cand}) are appended to the
    standardised records constructed in Section~\ref{subsec:completeness}. The final
    dataset is then exported into a lightweight SQLite database comprising five core
    tables (observatories, cameras, images, observations, and satellites), together
    with an auxiliary table \texttt{imgTotales} that stores the number of processed
    images per observing date.\\

    \begin{figure}[h!]
    \centering
    \includegraphics[width=\hsize]{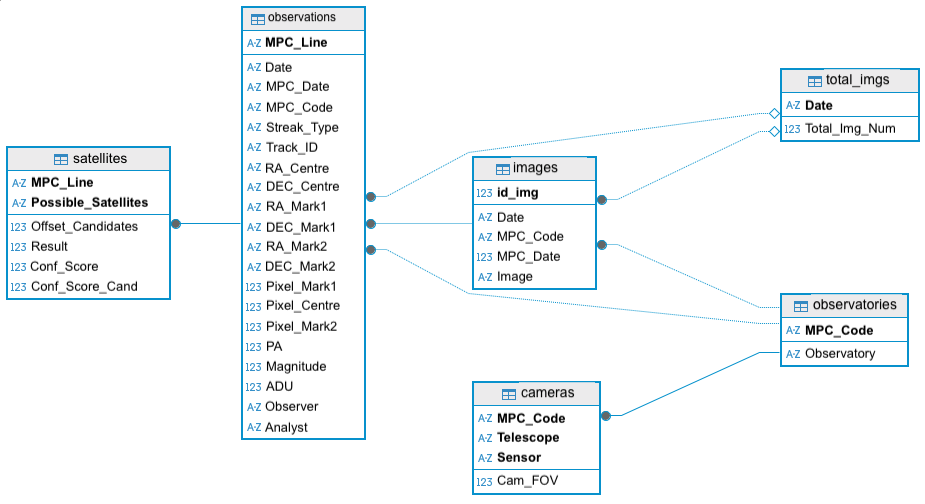}
    \caption{Database schema used for storing observation and object
    cross-identifications.}
    \label{fig:db_schema}
    \end{figure}

\section{Discussion}

\subsection{Comparison with manual inspection}

A direct comparison between StreakMind and traditional manual inspection highlights several practical advantages of the automated approach. First, the method is inherently scalable: the inference pipeline processes entire nights of observations in timescales of minutes, or even seconds for subsets of frames, whereas manual examination of the same dataset typically requires several hours of uninterrupted work. This difference becomes particularly relevant for observatories generating large nightly volumes of data, where exhaustive human inspection is no longer feasible.

In addition to its computational efficiency, the system exhibits a strong ability to detect faint streaks. Several of the detections recovered by StreakMind correspond to low S/N features that are difficult to identify through visual inspection alone. The automated processing therefore provides a more uniform sensitivity across the full dataset, independent of observer fatigue or subjective thresholds.

Another practical benefit is the immediate integration of results into the database structure described in Sect.~\ref{subsec:sat_xid_export}. The direct generation of standardised records and satellite cross-identifications removes the need for manual cataloguing tasks and ensures reproducible handling of both detections and astronomical objetcts identifications, reducing the overhead typically associated with cataloguing and post-processing.

Regarding geometric accuracy, the model performs robustly for streaks spanning up to approximately half of the image width. In this regime, both the detection rate and the OBB shape reconstruction remain highly reliable, making StreakMind a robust alternative to manual inspection. For very long streaks (exceeding $\sim 50\%$ of the image width), the detection remains stable, but the geometric adjustment requires a dedicated photometric-based post-processing step. This additional stage slightly reduces the positional precision compared to the short- and mid-length regime, although the streaks are still consistently identified and classified.

Overall, StreakMind provides a substantial improvement in efficiency, reproducibility, and sensitivity compared to manual inspection. It also maintains competitive geometric performance across most of the streak-length distribution encountered in this dataset.

\subsection{Perspectives for integration and future work}
\label{subsect:future_work}

The results presented in this work demonstrate the feasibility of an end-to-end automated pipeline for the detection, geometric characterisation, association, and identification of streaks produced by artificial satellites and space debris in ground-based astronomical images. All identifications discussed in this paper correspond exclusively to artificial objects, including operational satellites, rocket bodies and space debris, cross-matched against external satellite ephemerides.

Several extensions are currently under development to enhance both the scientific output and the operational usability of the pipeline. A primary forthcoming addition is a dedicated photometric analysis module, which will enable the estimation of the apparent brightness of the objects associated with each detected streak, along with S/N values that will allow for a more rigorous assessment of detection sensitivity. This functionality builds naturally upon the photometric information already exploited for geometric elongation and will allow for the inclusion of brightness measurements as first-class parameters in the detection database. In parallel, a graphical user interface is being developed to facilitate routine use of the pipeline by observers and analysts, reducing the barrier to adoption in operational environments and collaborative campaigns.

On the identification side, the current satellite cross-matching strategy will be extended through additional external services. In particular, an independent verification layer based on the SatChecker database\footnote{\url{https://satchecker.readthedocs.io/en/latest/}}, which provides orbital elements and metadata for a wide range of satellites, rocket bodies, and debris objects, is planned to further improve robustness and confidence assessment.

Beyond artificial objects, future versions of the pipeline will incorporate dedicated identification pathways for NEOs. Automated queries to the \texttt{Skybot} service \citep{Berthier_06} and to JPL's Horizons system \citep{Giorgini_96} will allow for the acquisition of ephemerides for known asteroids predicted to traverse the field of view. This extension will enable the confirmation of faint asteroid streaks that may be present in the images but remain undetected during the initial inference stage, as well as the automatic generation of MPC-compatible astrometric reports for submission to the MPC.

Finally, an important medium-term objective is the deployment of the detection model at additional ground-based observatories and the exploitation of larger archival datasets. Transfer learning offers a promising strategy in this context, allowing the model trained on La Sagra Observatory data to be rapidly adapted to new sites using comparatively small, site-specific datasets. As shown in this work, the performance of such models can be further enhanced through the controlled inclusion of synthetically generated streaks, particularly for underrepresented distributions. In the longer term, training the initial model on a more diverse, multi-observatory dataset may further improve generalisation and portability across instruments and observing conditions.

An important consideration for the broader applicability of the pipeline is its dependence on the specific observational setup, including telescope optics, camera characteristics, pixel scale, and typical seeing conditions. While the current model has been trained and validated exclusively on data from La Sagra Observatory, the underlying architecture is instrument-agnostic: all image pre-processing, OBB detection, and photometric elongation procedures operate on pixel-based representations and do not embed instrument-specific parameters.

Future deployment on other telescopes, including large-aperture instruments, will rely on transfer learning and site-specific fine-tuning. The adaptation strategy will depend on the characteristics of the new instrument, such as field of view, pixel scale, and noise properties. In cases of moderate differences, existing models can be adjusted using relatively small local datasets. If the new data distribution significantly departs from that of the original training set, new models may be trained from scratch. Synthetic streak augmentation, as demonstrated in this work, will further support adaptation by reproducing observational conditions not represented in the training data. Taken together, these approaches ensure that the pipeline can be extended to different observatories and camera configurations, facilitating integration into diverse operational environments while preserving detection performance and reliability.

\section{Conclusions}
\label{sec:conclusions}

We have introduced StreakMind, an AI-based pipeline designed for the automated detection and characterisation of linear streaks in ground-based astronomical images originally acquired for astrometric and photometric purposes. The system integrates a YOLO11 oriented bounding box detector trained on a stratified combination of real and synthetic images together with a geometric framework that converts FITS-based annotations into consistent PNG-space OBB labels. This approach enables robust operation across heterogeneous observing conditions and streak morphologies that span a wide range of lengths and signal-to-noise ratios while maintaining astrometric coherence throughout the pipeline.

Quantitative evaluation on an independent test set showed that the detector achieves high levels of precision and recall under a strict IoU threshold, with most real streaks successfully detected. Performance is particularly reliable for short- and intermediate-length streaks, for which both the detection rate and the geometric reconstruction of the OBBs remain stable. For very long streaks, detection remains robust, while geometric refinement benefits from the dedicated photometric post-processing stage. The bright-star filtering procedure proved effective during the subsequent inference stage by suppressing the diffraction spikes that dominate the false-positive detections. These results are reflected in the final performance metrics on the test set, where the model reached a precision of 94\,\% and a recall of 97\,\% under a strict IoU threshold.

A further advantage of the pipeline lies in the direct integration of the inference results into a normalised detection database. Each detection is transformed into an MPC-style record that is enriched with geometric and temporal information and subsequently cross-identified with external satellite ephemerides using a sequential offset filtering and confidence evaluation process. This produces a structured and reproducible dataset suitable for large-scale statistical analyses of space-object contamination and for the systematic exploitation of archival observations.

Overall, StreakMind represents a first step towards efficient analyses of orbital-object contamination. Moreover, we have provided a foundation for the integration of StreakMind into broader astrometric workflows.

\begin{acknowledgements}
This work was carried out in collaboration with the Instituto de Astrofísica de Andalucía (IAA–CSIC), which operates the La Sagra Observatory.

René Duffard and Nicolas Morales acknowledge financial support from the Severo Ochoa grant CEX2021-001131-S funded by MCIN/AEI/10.13039/501100011033. Google Colab was employed for part of the model training and experimentation. This work has made use of data from the European Space Agency (ESA) mission
\emph{Gaia} (\url{https://www.cosmos.esa.int/gaia}), processed by the Gaia
Data Processing and Analysis Consortium (DPAC,
\url{https://www.cosmos.esa.int/web/gaia/dpac/consortium}). Funding for the DPAC
has been provided by national institutions, in particular the institutions
participating in the Gaia Multilateral Agreement.
\end{acknowledgements}

\bibliographystyle{bibtex/aa} 
\bibliography{bibtex/biblio} 

\begin{appendix}
\section{OBB construction and clipping procedure}
\label{app:obb}

We first computed the directional vector and its length:
\[
\vec{d} = (x_2 - x_1,\; y_2 - y_1), 
\qquad
L = \|\vec{d}\| = \sqrt{(x_2 - x_1)^2 + (y_2 - y_1)^2}.
\]
To provide contextual background, the streak was extended by a fixed margin \(m = 4\)\,px along its axis, yielding a nominal half-length:
\[
\frac{L'}{2} = \frac{L + 2m}{2} = \frac{L}{2} + m.
\]
The OBB is defined with a constant width \(W = 16\)\,px, corresponding to a symmetric expansion of \(\pm W/2\) pixels along the direction orthogonal to the streak.

We define a local coordinate system \((d_\ell, d_w)\), where \(d_\ell\) is aligned with the streak axis (longitudinal component), and \(d_w\) is perpendicular (width direction). The initial (unrotated) corner positions in this frame are
\[
V_{\rm local} = \left\{ \left(-\tfrac{L'}{2}, -\tfrac{W}{2}\right),\; \left(\tfrac{L'}{2}, -\tfrac{W}{2}\right),\;
\left(\tfrac{L'}{2}, \tfrac{W}{2}\right),\; \left(-\tfrac{L'}{2}, \tfrac{W}{2}\right) \right\}.
\]

\medskip
\noindent The vertex sequence defined in $V_{\rm local}$ follows a counter-clockwise order starting from the bottom-left corner in the local frame. However, after rotating and translating these coordinates into the PNG image reference system (which includes a vertical flip), the vertices are explicitly reordered to match the clockwise format expected by object detection frameworks. This reordering does not alter the geometry of the box and is described in detail below.

The angle of the streak was computed as
\[
\theta = \atanTwo(y_2 - y_1,\; x_2 - x_1),
\]
and the centre of the OBB (after margin extension) is
\[
C_x = \frac{x_1 + x_2}{2} + m \cos\theta, \qquad C_y = \frac{y_1 + y_2}{2} + m \sin\theta.
\]

Before rotation, we applied a symmetric clamping to the half-width of the OBB to ensure it remains within vertical bounds:
\[
\frac{W_{\text{clamped}}}{2} = \min\left( \frac{W}{2},\; C_y,\; H_{\text{img}} - C_y \right).
\]
The clamping is performed before the longitudinal adjustment to guarantee that, after axis-aligned scaling is applied, the vertical extent of the box remains fully contained within the image bounds even in cases where the centre lies close to the edges.

Each of the four vertices in \(V_{\rm local}\) is then rotated by the angle
\(\theta\) and translated to the centre \((C_x, C_y)\), yielding their global
coordinates in the PNG reference frame:
\[
\begin{pmatrix}
x_g^{(k)} \\[2pt]
y_g^{(k)}
\end{pmatrix}
=
\begin{pmatrix}
C_x \\[2pt]
C_y
\end{pmatrix}
+
\begin{pmatrix}
\cos\theta & -\sin\theta \\
\sin\theta &  \cos\theta
\end{pmatrix}
\begin{pmatrix}
d_{\ell}^{(k)} \\[2pt]
d_{w}^{(k)}
\end{pmatrix},
\qquad k = 1,\dots,4.
\]

To keep all box vertices within the image frame along the streak direction,
\[
x_g \in [0, W_{\text{img}}),\qquad y_g \in [0, H_{\text{img}}),
\]
we evaluated each vertex and computed a positive scaling factor \(f\) (applied to \(d_\ell\)) whenever a corner lies out of bounds:
\begin{align}
\text{if } x_g < 0:\quad
f &= \frac{-C_x + \sin\theta\,d_w}{\cos\theta\,d_\ell},\\
\text{if } x_g > W_{\text{img}}:\quad
f &= \frac{W_{\text{img}} - C_x + \sin\theta\,d_w}{\cos\theta\,d_\ell},\\
\text{if } y_g < 0:\quad
f &= \frac{-C_y - \cos\theta\,d_w}{\sin\theta\,d_\ell},\\
\text{if } y_g > H_{\text{img}}:\quad
f &= \frac{H_{\text{img}} - C_y - \cos\theta\,d_w}{\sin\theta\,d_\ell}.
\end{align}

The final scaling factors for the two halves of the box are defined as
\[
f_{\rm back} = \min\left(f \text{ for all corners with } d_\ell < 0\right),
\]
\[
f_{\rm forward} = \min\left(f \text{ for all corners with } d_\ell > 0\right),
\]
with both saturated to the range \([0, 1]\).

We then computed the corrected half-lengths:
\[
\ell_{\rm back} = \tfrac{L'}{2}\cdot f_{\rm back}, 
\qquad 
\ell_{\rm forward} = \tfrac{L'}{2}\cdot f_{\rm forward},
\]
and we redefined the OBB corners accordingly,
\begin{equation}
\begin{aligned}
V_{\rm local}^{\rm clipped} = \Big\{ 
&\left(-\ell_{\rm back},\; -\tfrac{W_{\text{clamped}}}{2}\right),\quad 
\left(\ell_{\rm forward},\; -\tfrac{W_{\text{clamped}}}{2}\right),\\
&\left(\ell_{\rm forward},\; \tfrac{W_{\text{clamped}}}{2}\right),\quad 
\left(-\ell_{\rm back},\; \tfrac{W_{\text{clamped}}}{2}\right) 
\Big\}
\end{aligned}.
\label{eq:VlocalClipped}
\end{equation}

This asymmetric scaling ensures that if one end of the box exceeds the image boundaries, only that portion is trimmed, preserving the remainder of the annotation and maintaining geometrical consistency.

\medskip
\noindent The order of vertices in \(V_{\rm local}^{\rm clipped}\) follows the same convention as in \(V_{\rm local}\), preserving the counter-clockwise sequence in the local frame. However, this order is later adjusted once the corners are projected into global coordinates, using the same reordering strategy previously described to ensure consistency in the PNG reference frame.

\medskip
\noindent \paragraph{Vertex ordering for annotation.}  
The final ordering of the OBB corners is explicitly enforced to match the format required by YOLO-based detection frameworks. This format expects a clockwise sequence of vertices starting from the top-left corner of the image.

To ensure this, the four projected vertices are analysed in the PNG coordinate system (with top-left origin and vertical flip). The procedure is as follows:

\begin{itemize}
  \item Select the two vertices with the smallest $x$ coordinates (i.e. the leftmost).
  \item Among these vertices, the one with the smallest $y$ is labelled as `top left' (TL), and the other is labelled as `bottom left' (BL).
  \item The two remaining vertices (with largest $x$ values) are labelled analogously. The smallest $y$ becomes `top right' (TR), and the other becomes `bottom right' (BR).
\end{itemize}

This process produces the final ordered set:
\[
V_{\rm ordered} = \{ \text{TL},\ \text{TR},\ \text{BR},\ \text{BL} \}.
\]

\end{appendix}

\end{document}